\documentclass[journal]{IEEEtran}
\usepackage[T1]{fontenc}
\usepackage{amsmath,amssymb}
\usepackage{booktabs,array,tabularx}
\usepackage{graphicx}
\usepackage{cite}
\usepackage[hyphens]{url}
\usepackage{xurl}
\usepackage[hidelinks]{hyperref}
\hypersetup{pdftitle={Fast Reassessment of Prolonged Electricity Shortages under Shared Fuel Interruptions},pdfauthor={Dogan Urgun}}
\newcommand{\E}{\mathbb{E}}
\newcommand{\Prb}{\mathbb{P}}
\newcommand{\ind}{\mathbf{1}}
\newcommand{\Qd}{\mathcal{Q}_d}
\newtheorem{proposition}{Proposition}
\makeatletter
\def\@begintheorem#1#2{\@IEEEtmpitemindent\itemindent\relax\topsep 0pt\rmfamily\trivlist%
  \item[]\textit{\indent #1\ #2.} \itemindent\@IEEEtmpitemindent\relax}
\def\@opargbegintheorem#1#2#3{\@IEEEtmpitemindent\itemindent\relax\topsep 0pt\rmfamily\trivlist%
  \item[]\textit{\indent #1\ #2\ (#3).} \itemindent\@IEEEtmpitemindent\relax}
\makeatother
\begin{document}
\title{Fast Reassessment of Prolonged Electricity Shortages under Shared Fuel Interruptions}
\author{Dogan~Urgun,~\IEEEmembership{Member,~IEEE}%
\thanks{This work has been submitted to the IEEE for possible publication. Copyright may be transferred without notice, after which this version may no longer be accessible.}\thanks{D. Urgun is with the Department of Electrical and Electronics Engineering, Karab\"uk University, 78050 Karab\"uk, T\"urkiye (e-mail: durgun@karabuk.edu.tr).}}
\maketitle

\begin{abstract}
When several generators share a fuel supply, one interruption can disable them together and cause electricity shortages lasting days. Such interruptions can be made less frequent, for example by weatherizing gas supply, or shorter, for example by on-site backup fuel. Yet traditional resource adequacy indices, expected unserved energy and loss-of-load hours, fail to distinguish such measures whenever they remove the same expected time without fuel, because for interruptions independent of system conditions these indices depend on the interruptions only through this time. Their difference is revealed by prolonged-shortage exposure, the expected annual hours in shortages of at least a given duration, whose exact chronological calculation must be repeated for every interruption scenario. To reduce this cost, a fast reassessment method with guaranteed accuracy is proposed. Network calculations are performed once and condensed into a restoration response, the additional prolonged-shortage hours caused by one interruption of each duration, from which any scenario is evaluated as a weighted sum without new simulation. An hour-by-hour error bound determines whether this evaluation is accurate enough, and exact dynamic programming is applied otherwise, so that every result lies within a requested tolerance of the exact value on the same Monte Carlo samples. On the Reliability Test System of the Grid Modernization Laboratory Consortium, most evaluations were accepted, fully accepted scenarios typically ran more than 200 times faster than exact calculation, and total computation time decreased in every workload. Measures against fuel interruptions can thus be screened quickly and accurately by their effect on prolonged shortages.
\end{abstract}

\begin{IEEEkeywords}
Dynamic programming, fuel supply, Monte Carlo methods, power system reliability, resource adequacy.
\end{IEEEkeywords}

\section*{Nomenclature}
\begin{IEEEdescription}[\IEEEusemathlabelsep\IEEEsetlabelwidth{RTS-GMLC}]
\item[DP] Dynamic programming.
\item[EUE] Expected unserved energy.
\item[LOLH] Loss-of-load hours.
\item[MCS] Monte Carlo simulation.
\item[RMSE] Root mean square error.
\item[RTS-GMLC] Reliability Test System of the Grid Modernization Laboratory Consortium.
\item[SE] Standard error.
\item[$A,B,N$] Hours always, conditionally, and never short.
\item[$\alpha_t^\pm$] Bounds on the rate of incidents that alone make $t$ qualify.
\item[$b_d(s)$] Span coefficient for span $s$ and threshold $d$.
\item[$c_d$] Reward of an ordinary history.
\item[$d$] Prolonged-shortage threshold (h).
\item[$D,P_D$] Restoration time (h) and its distribution.
\item[$F_d$] First-order estimate of $I_d$ (weighted sum).
\item[$\varphi_1,\varphi_2$] Expected Poisson count beyond the first, and probability of two or more.
\item[$\mathcal G$] Group of generators sharing a fuel supply.
\item[$H_d$] Shortage hours in runs of at least $d$ hours.
\item[$I_d$] Fuel-induced increment of the expected reward.
\item[$\kappa_d$] Absolute bound on the averaged first-order error.
\item[$\ell_t^0,\ell_t^1$] Shedding with $\mathcal G$ available and removed.
\item[$\lambda,\rho$] Rate of incidents (shared fuel interruptions) per year and per hour.
\item[$\mu_d$] Prolonged-shortage exposure (h/year).
\item[$\nu_t$] Mean number of incidents that can affect hour $t$.
\item[$\omega$] Ordinary history of equipment availability.
\item[$\pi$] Probability that an hour is covered.
\item[$\Psi_d(u)$] Restoration response for duration $u$.
\item[$\Qd$] Hours that incidents can make qualify.
\item[$R_d,R_d^\pm$] Remainder $I_d-F_d$ and its local bounds.
\item[$s$] Span from first to last required hour.
\item[$\tau_{\mathrm a},\tau_{\mathrm r}$] Absolute and relative tolerances.
\item[$\theta$] Interruption scenario $(\rho,P_D)$.
\item[$U(s)$] Expected restoration time beyond $s$ hours.
\item[$z_t,m_t,o_t$] Timing summaries of hour $t$.
\item[$Z_t,\chi_t$] Shortage and coverage indicators.
\end{IEEEdescription}

\section{Introduction}\label{sec:introduction}
\IEEEPARstart{W}{hen} a shared fuel supply is interrupted, several electricity generators can become unavailable at the same time. During Winter Storm Elliott in December 2022, about 18\% of the anticipated generating resources in the U.S. portion of the Eastern Interconnection, including units already out of service, were unavailable at the worst point, and fuel issues, mostly related to natural gas, accounted for 24\% of the megawatts lost to unplanned outages, derates, and failures to start. Firm load had to be shed, leaving hundreds of thousands of customers without power for several hours in extreme cold~\cite{ferc2023}. Fuel supply risks continue to be identified in reliability assessments~\cite{nerc2025,iea2025}, and in January 2025, unplanned generator outages again reached about a tenth of peak demand, although no load was shed~\cite{ferc2025}. Generator outages of comparable magnitude recurred during Winter Storms Fern and Gianna in early 2026~\cite{ferc2026}. The consequences of shared fuel interruptions can be reduced by measures that make interruptions less frequent, such as weatherization of gas production and pipeline facilities or firm transportation contracts, or by measures that make them shorter, such as on-site backup fuel or faster restoration of supply. Dual-fuel capability, for example, helped maintain reliability during natural gas constraints in January 2025~\cite{ferc2025}. For these measures to be compared, the duration of the resulting shortages must be quantified.

\begin{figure*}[!t]
\centering
\includegraphics[width=0.68\textwidth]{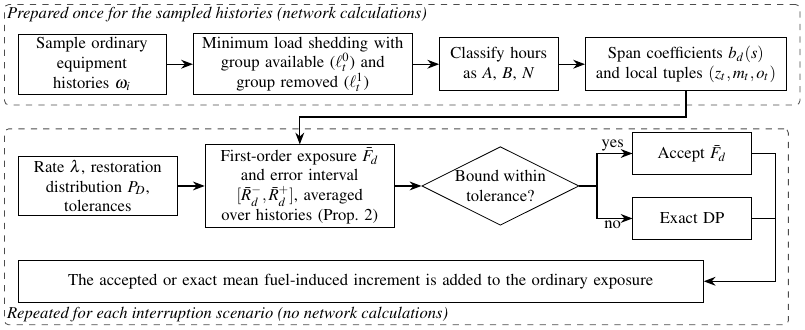}
\caption{Overview of the method. Network calculations are performed once, and each interruption scenario is evaluated by the weighted sum or, where the bound is inconclusive, by exact DP.}
\label{fig:overview}
\end{figure*}

This information is not provided by expected unserved energy (EUE) or loss-of-load hours (LOLH). For interruptions independent of system conditions, both indices depend on the interruptions only through the expected time without fuel, so that they cannot distinguish measures that reduce this time equally. They are also annual totals, and the same totals can be accumulated through isolated shortage hours or through one shortage lasting several days. Duration-sensitive metrics have therefore been recommended~\cite{esig2024,stenclik2023,dent2025}, and multiday fuel and weather stresses have been examined in regional adequacy studies~\cite{pjm2018fuel,isone2023peat}. In this paper, prolonged-shortage exposure is measured by the expected annual hours that belong to shortages lasting at least a selected duration, such as 24 or 48~h. Mitigation measures with equal EUE and LOLH can differ substantially in this exposure, as shown in Section~\ref{sec:results}.

This exposure cannot be inferred from interruption durations alone. During a fuel interruption, demand may still be met by the remaining system in some hours, so that the shortage ends before fuel is restored. Conversely, a shortage already caused by other equipment failures may be extended by a short interruption. Exposure is therefore determined by the coincidence of interruptions with the sequence of operating conditions, and it must be evaluated chronologically.

Such an evaluation is provided by sequential Monte Carlo simulation (MCS), in which equipment histories are sampled and the ability to meet demand is checked hour by hour~\cite{roman1992,peng2019,singh2019}. Fuel interruptions are rare, so an interruption is sampled in only a few simulated years, and many years must be simulated for an accurate estimate. This sampling variation can be removed by conditional Monte Carlo, in which the sampled interruptions are replaced by their expected effect on each equipment history~\cite{rubinstein2016,mello1997}. The expectation can be computed exactly by dynamic programming (DP), in which active interruptions and the current shortage length are tracked hour by hour through finite Markov-chain imbedding~\cite{fu1996}.

The DP, however, must be repeated over all sampled histories for every interruption scenario, and a comparison of mitigation alternatives involves many scenarios. Network results have been stored for reuse~\cite{peng2019,peng2021,xu2025}, but not in a form that captures whether hours separated in time are interrupted together.

In this paper, a reassessment method is proposed in which the exact calculation is replaced by a weighted sum whenever a bound shows that the sum is accurate enough (Fig.~\ref{fig:overview}). For each sampled history, load shedding is calculated once with the fuel-dependent generators available and once with them removed. From the two results, a \emph{restoration response} is derived, whose weighted sum approximates the fuel-induced exposure under any interruption scenario. The error of this approximation, caused by closely spaced interruptions, is bounded from the hours near each affected hour, and exact DP is performed only where the bound exceeds the tolerance.

The main contributions of this paper are as follows.
\begin{enumerate}
\setlength{\itemsep}{0pt}\setlength{\parskip}{0pt}
\item A restoration response is introduced, from which prolonged-shortage exposure under any interruption frequency and restoration-time distribution is approximated by a weighted sum with a bounded error, while network calculations are performed only once.
\item A local error bound is derived. Because only hours near each affected hour are considered, it remains informative where a block bound over entire shortage periods fails.
\item An error-controlled reassessment procedure is formulated, which accepts the weighted sum when the bound meets the tolerance and applies exact DP otherwise.
\item The exact expected exposure at a threshold is shown to depend only on the mean restoration time and the probabilities of restoration times shorter than the threshold, which identifies the restoration data to be collected.
\item A planning comparison shows that measures with identical EUE and LOLH can differ substantially in prolonged-shortage exposure.
\end{enumerate}
To support reproducibility, the code, the sampled histories, and the scripts used to produce the results are publicly available at {\renewcommand{\UrlBreaks}{\do\/\do\_}\url{https://github.com/DoganUrgun/Fast_Reassessment_Prolonged_Shortages}}, and the benchmark data are publicly available as described in Section~\ref{sec:implementation}.

The remainder of this paper is organized as follows. Related work is reviewed in Section~\ref{sec:related}, and the problem is defined in Section~\ref{sec:problem}. The method is presented in Section~\ref{sec:method}, implementation details are given in Section~\ref{sec:implementation}, and the results are reported in Section~\ref{sec:results}. Conclusions are drawn in Section~\ref{sec:conclusion}, and proofs, constructions, and an example are given in the appendices.

\section{Related Work}\label{sec:related}
Frequency and duration indices of composite systems have been evaluated through conditional probabilities~\cite{melo1993}, chronological interruption sequences have been reconstructed by pseudo-sequential simulation~\cite{mello1997}, and probability distributions of interruption duration have been obtained analytically and by MCS~\cite{silva2005}. In these studies, the system model is fixed, and shared-interruption assumptions are not examined.

The computational effort of chronological evaluation has been reduced by rare-event sampling~\cite{kiel2023}, importance resampling~\cite{bjerkebaek2024}, and cross-entropy methods~\cite{manso2024}. Selected uncertainties have also been evaluated without sampling, by sequential analytical methods~\cite{ge2025} and by finite Markov-chain imbedding of run statistics~\cite{fu1994,fu1996,fu2002}. These approaches are complementary to the proposed method, in which shared-interruption uncertainty is integrated analytically on sampled histories.

Repeated evaluation has been accelerated by retaining network results. Configuration coefficients and failure-order contributions have been retained for changes in component reliability parameters~\cite{peng2019} and for hardening analysis~\cite{peng2021}. Contingency-conditioned load-shedding functions have been developed for changed load and renewable conditions~\cite{xu2025}, and repeated optimization has been reduced by shadow-price and impact-increment methods~\cite{hou2021} and by classifiers that identify system and bus failures without optimal power flow calculations, alone~\cite{urgun2019} or combined with importance sampling~\cite{urgun2020}. These approaches accelerate the evaluation of individual system states, from which expectation indices such as EUE and LOLH are obtained, but the joint interruption of hours separated in time, which determines prolonged-shortage exposure, is not represented in them.

The interruption model used here is a Poisson process of random intervals, whose coverage has been studied in~\cite{shepp1972}, and first-order expansions in the intensity of a Poisson process are provided by perturbation analysis~\cite{last2017}. Dependency neighborhoods are used in the method of Chen and Stein to bound the error of Poisson approximations~\cite{arratia1989}, whereas here they bound the error of a first-order expansion of the prolonged-shortage reward.

Physical coupling between electricity and gas systems~\cite{zeng2020,li2022}, temperature-dependent failures~\cite{murphy2019}, correlated outages~\cite{gunda2025}, and time-coupled operation~\cite{leibowicz2024} have been represented in broader reliability models. Here, interruptions are prescribed independently of equipment histories, which isolates restoration effects and allows the paired network histories to be reused.

\section{Problem Definition}\label{sec:problem}
\subsection{System and Interruption Model}
The study system is described by a transmission network, hourly demand and renewable-availability profiles, failure and repair models for generators and branches, and a group $\mathcal G$ of generators supplied from a common fuel source. Equipment availability over the hours $t=0,\ldots,T-1$, with $T=8760$ for an annual study, is sampled from the failure and repair models. Each sampled chronology $\omega$, which excludes shared fuel interruptions, is termed an \emph{ordinary history}. In each hour, load shedding is minimized subject to the available generation, dc network balance, and line limits. No storage, commitment, or other operating state is carried between hours, so that the load shedding in each hour depends only on the equipment availability and profiles of that hour. With $\mathcal G$ removed as a whole, two network calculations per hour therefore suffice (Section~\ref{sec:paired}).

\begin{figure*}[!t]
\centering
\includegraphics[width=0.94\textwidth]{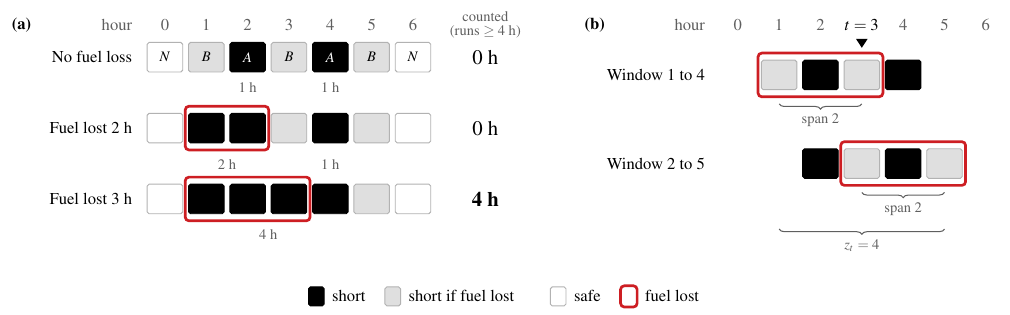}
\caption{Running example with threshold $d=4$~h. (a) Hour classes $A$, $B$, and $N$ and the shortage caused by a 2-h and a 3-h fuel loss. Only the 3-h loss joins hours 1 to 4 into a run that is counted in $H_4$. (b) For hour $t=3$, the two eligible 4-h windows, their required hours (gray), and the shortest fuel loss that makes each window entirely short. Both spans are 2, and $z_t=4$, $m_t=2$, and $o_t=0$.}
\label{fig:classes}
\end{figure*}

Shared fuel interruptions are modeled as \emph{incidents}. Incident starts are modeled by independent Poisson counts at each integer hour with mean $\rho=\lambda/8760$, where $\lambda$ is the expected number of incidents per year. A bounded positive-integer restoration time $D$, distributed according to $P_D$, is assigned independently to each incident and gives its duration in hours. During hours $v,\ldots,v+D-1$ of an incident started at hour $v$, every generator in $\mathcal G$ is removed, in addition to any outages recorded in $\omega$. Overlapping incidents are allowed, and incidents started before $t=0$ are included, so that every hour of the year is covered with the same probability. Coverage of hour $t$ by at least one incident is indicated by $\chi_t$. The incident process is assumed to be independent of $\omega$, and an \emph{interruption scenario} is denoted by $\theta=(\rho,P_D)$.

\subsection{Prolonged-Shortage Exposure}
The minimum load shedding at hour $t$, determined by $\omega$ and $\chi_t$, is denoted by $\ell_t$. A shortage is indicated by $Z_t=\ind\{\ell_t>\varepsilon\}$, where $\varepsilon$ is a small shedding tolerance, and a \emph{run} is a maximal sequence of consecutive shortage hours. For a duration threshold $d$, the annual prolonged-shortage reward is defined as
\begin{equation}
H_d(Z)=\sum_{r\in\mathcal R(Z)}r\,\ind\{r\ge d\},
\label{eq:reward}
\end{equation}
where $\mathcal R(Z)$ is the multiset of run lengths, with runs clipped at the year boundaries. The reward $H_d$ is the number of shortage hours in runs of at least $d$ hours, so that a 30-h shortage is counted as 30~h in $H_{24}$ and not counted in $H_{48}$. In run theory, $H_d$ is known as the number of successes in success runs of length at least $d$~\cite{fu2002}. The prolonged-shortage exposure is defined as $\mu_d(\theta)=\E H_d$, with the expectation taken over both ordinary histories and incidents.

\subsection{Ordinary and Fuel-Induced Components}
The reward obtained in the absence of incidents is denoted by $c_d(\omega)$, and the fuel-induced increment is defined as
\begin{equation}
I_d(\omega;\theta)=\E_\theta[H_d\mid\omega]-c_d(\omega),
\label{eq:increment}
\end{equation}
where $\E_\theta$ denotes expectation over incidents under scenario $\theta$ for a fixed ordinary history. The exposure is thereby decomposed as
\begin{equation}
\mu_d(\theta)=\E_\omega c_d(\omega)+\E_\omega I_d(\omega;\theta).
\label{eq:target}
\end{equation}
The ordinary component does not depend on $\theta$. It is estimated once and cancels when interruption scenarios are compared. The problem is thus to evaluate $\E_\omega I_d(\omega;\theta)$ for many scenarios $\theta$ on the same sampled ordinary histories, with network calculations performed only once.

\section{Methodology}\label{sec:method}
The method proceeds in four steps (Fig.~\ref{fig:overview}), namely hour classification, a first-order estimate, an error bound, and acceptance or exact DP (Sections~\ref{sec:paired} to~\ref{sec:reassessment}). Restoration information and estimation follow in Sections~\ref{sec:information} and~\ref{sec:estimation}.

\subsection{Paired Loss Histories and Hour Classes}\label{sec:paired}
For each ordinary history, the minimum load shedding is computed twice, once with $\mathcal G$ at its ordinary availability, giving $\ell_t^0$, and once with $\mathcal G$ removed, giving $\ell_t^1\ge\ell_t^0$. Because $\mathcal G$ is removed as a whole and the hours are dispatched independently, the shedding under any incident realization is
\begin{equation}
\ell_t=\ell_t^0+\chi_t(\ell_t^1-\ell_t^0).
\label{eq:paired}
\end{equation}
Equation~\eqref{eq:paired} also shows why EUE and LOLH cannot distinguish restoration alternatives with equal $\rho\E D$. Each hour is covered with the stationary probability $\pi=1-e^{-\rho\E D}$, so that both indices depend on $\theta$ only through $\rho\E D$, and equal-mean restoration-time distributions yield equal EUE and LOLH at the same rate. Prolonged-shortage exposure, in contrast, depends on which hours are covered together. To describe this timing, each hour is assigned to one of three classes,
\begin{equation}
\begin{aligned}
A&=\{t:\ell_t^0>\varepsilon\},\qquad N=\{t:\ell_t^1\le\varepsilon\},\\
B&=\{t:\ell_t^0\le\varepsilon<\ell_t^1\}.
\end{aligned}
\label{eq:classes}
\end{equation}
Shedding occurs in class $A$ regardless of fuel availability and in class $B$ only when an incident is active, whereas demand is met in class $N$ even with $\mathcal G$ removed. A window of consecutive hours is therefore entirely short if and only if it contains no class-$N$ hour and all of its class-$B$ hours, termed its \emph{required hours}, are covered. The ordinary reward is $c_d=H_d(\ind_A)$.

The chronology in Fig.~\ref{fig:classes}(a) is used as a running example with $d=4$. A 2-h incident on hours 1 and 2 leaves runs of only 2~h and 1~h, whereas a 3-h incident on hours 1 to 3 creates a 4-h run over hours 1 to 4, which adds 4~h to $H_4$.

\subsection{Restoration Response}\label{sec:response}
The restoration response gives the additional prolonged-shortage hours caused by a single incident as a function of its duration. A direct evaluation is complicated by the fact that the qualification of a run depends on its total length. The reward is therefore expressed through windows, whose shortage depends only on the coverage of their own required hours. With $W_k$ denoting the number of entirely short windows of $k$ hours, a run of length $r$ contains $(r-k+1)_+$ such windows, where $(x)_+=\max(x,0)$. Hence
\begin{equation}
H_d=dW_d-(d-1)W_{d+1},
\label{eq:window}
\end{equation}
which assigns zero to runs shorter than $d$ and exactly $r$ to each run of length $r\ge d$. A 30-h run, for example, contains seven windows of 24~h and six of 25~h, and $24\cdot7-23\cdot6=30$.

A single incident of duration $u$ covers all required hours of a window if it starts no later than the first required hour $f$ and ends no earlier than the last required hour $l$. This occurs for $(u-s)_+$ start hours, where $s=l-f$ is termed the \emph{span} of the window. A window is termed \emph{eligible} if it contains no class-$N$ hour and at least one required hour. With $\eta_k(s)$ denoting the number of eligible windows of $k$ hours and span $s$, the integer span coefficients are defined as
\begin{equation}
b_d(s)=d\,\eta_d(s)-(d-1)\,\eta_{d+1}(s),\qquad 0\le s\le d.
\label{eq:coefficients}
\end{equation}

\begin{proposition}[Restoration response]\label{prop:response}
With $Z^{v,u}$ denoting the shortage sequence produced by a single incident of duration $u$ started at hour $v$, the additional qualifying hours, summed over all starts that can affect the year, are
\begin{equation}
\Psi_d(u)=\sum_{v=1-u}^{T-1}\big[H_d(Z^{v,u})-c_d\big]
=\sum_{s=0}^{d}b_d(s)\,(u-s)_+.
\label{eq:response}
\end{equation}
\end{proposition}
The proof is given in Appendix~\ref{app:response}.

The response is nonnegative, nondecreasing, and piecewise linear in $u$, with breakpoints at the spans and a linear form for $u\ge d$. In the running example, the eligible four-hour windows cover hours 1 to 4 and 2 to 5, both with span 2 (Fig.~\ref{fig:classes}(b)), and the only eligible five-hour window, covering hours 1 to 5, has span 4. Hence $b_4(2)=8$, $b_4(4)=-3$, and $\Psi_4(u)=8(u-2)_+-3(u-4)_+$. The negative term applies when one incident covers all three required hours. A single five-hour run is then formed, which would otherwise be counted twice through the two four-hour windows.

The expected sum of single-incident effects is obtained by averaging the response over $D$ and multiplying by the start rate,
\begin{equation}
F_d(\theta)=\rho\,\E\Psi_d(D)=\rho\sum_{s=0}^{d}b_d(s)\,U(s),
\label{eq:firstorder}
\end{equation}
where $U(s)=\E(D-s)_+$ is the expected part of the restoration time beyond $s$ hours. A new scenario is therefore represented by the rate $\rho$ and the $d+1$ numbers $U(0),\ldots,U(d)$ alone. By the perturbation formula for Poisson functionals~\cite{last2017}, $F_d$ is the first-order term of $I_d$ in the incident intensity, and it is exact if no two incidents affect the same shortage. The remainder $R_d=I_d-F_d$ is caused by incidents that occur close together. It is positive when a prolonged shortage is created jointly by several incidents and negative when the same hours are added by more than one incident.

\subsection{Local Error Bound}\label{sec:bound}
The replacement of $I_d$ by $F_d$ is admissible only if $R_d$ is small, and $R_d$ must therefore be bounded without evaluating $I_d$. A simple bound $\kappa_d^{\mathrm{blk}}$ is obtained over blocks of consecutive non-$N$ hours from the number of incidents that intersect each block (Appendix~\ref{app:block}). Since this block bound counts interactions over entire blocks, its upper estimate grows with the cube of the block length (Appendix~\ref{app:block}), and it becomes uninformative in stressed systems, in which blocks extend over hundreds of hours.

A tighter bound is obtained hour by hour. An hour $t$ belongs to a run of at least $d$ hours if and only if it belongs to an entirely short window of length $d$, and all such windows lie within $[t-d+1,t+d-1]$. Hour $t$ can therefore be affected only by incidents intersecting this neighborhood, and the first-order sum is exact for $t$ unless two or more of them occur.

Only hours that qualify when every class-$B$ hour is covered, but not in the ordinary history, can be affected, and their set is denoted by $\Qd$. For $t\in\Qd$, the eligible $d$-hour windows containing $t$ are ordered by their start, and the first and last required hours of window $w$ are denoted by $f_w$ and $l_w$. The timing requirements around $t$ are summarized by three integers (Fig.~\ref{fig:classes}(b)),
\begin{equation}
\begin{aligned}
z_t&=l_{\mathrm{last}}-f_{\mathrm{first}},\qquad m_t=\min\nolimits_w(l_w-f_w),\\
o_t&=l_{\mathrm{first}}-f_{\mathrm{last}}.
\end{aligned}
\label{eq:geometry}
\end{equation}
The coverage of the hours from $f_{\mathrm{first}}$ to $l_{\mathrm{last}}$, a stretch with span $z_t$, determines whether $t$ qualifies. An incident of duration $u\le m_t$ cannot make $t$ qualify on its own, and an incident of duration $u\ge d$ can do so from exactly $u-o_t$ start hours, where $o_t$ may be negative.

\begin{proposition}[Local error enclosure]\label{prop:local}
Let $\varphi_1(x)=x-1+e^{-x}$ and $\varphi_2(x)=1-(1+x)e^{-x}$, and for $t\in\Qd$ let $\nu_t=\rho(\E D+z_t)$ and
\begin{equation}
\begin{aligned}
\alpha_t^-&=\rho\big[U(m_t)+(m_t-o_t)\Prb(D\ge d)\big],\\
\alpha_t^+&=\rho\big[U(m_t)+(m_t-o_t)\Prb(D>m_t)\big].
\end{aligned}
\label{eq:alpha}
\end{equation}
Then, for every ordinary history, $R_d^-\le R_d\le R_d^+$, where
\begin{align}
R_d^-&=-\sum_{t\in\Qd}\varphi_1(\alpha_t^+),\nonumber\\
R_d^+&=\sum_{t\in\Qd}\big[\varphi_2(\nu_t)-\alpha_t^-(1-e^{-\nu_t})\big],
\label{eq:localbound}
\end{align}
and
\begin{equation}
|R_d|\le\frac{\rho^2}{2}\sum_{t\in\Qd}(\E D+z_t)^2
\le\frac{|\Qd|\rho^2}{2}(\E D+2d-2)^2 .
\label{eq:scaling}
\end{equation}
\end{proposition}

The derivation is given in Appendix~\ref{app:local}. No independence between hours is assumed, and remainders of either sign are admitted. By~\eqref{eq:scaling}, the local bound is quadratic in the incident rate and, for a fixed threshold and restoration distribution, linear in the number of potentially affected hours. In the implementation, the local interval is also intersected with $[-\kappa_d^{\mathrm{blk}},\kappa_d^{\mathrm{blk}}]$ and with the range $[-F_d,|\Qd|-F_d]$ implied by $0\le I_d\le|\Qd|$.

\subsection{Error-Controlled Reassessment}\label{sec:reassessment}
For each scenario and threshold, the bounds are averaged over the $n$ sampled histories. With bars denoting sample means, the absolute error bound $\kappa_d=\max\{|\bar R_d^-|,|\bar R_d^+|\}$ and the lower estimate $\underline I_d=\max\{\bar F_d+\bar R_d^-,0\}$ are formed, and, for requested absolute and relative tolerances $\tau_{\mathrm a}$ and $\tau_{\mathrm r}$, the weighted sum is accepted if
\begin{equation}
\kappa_d\le\tau_{\mathrm a}+\tau_{\mathrm r}\,\underline I_d .
\label{eq:acceptance}
\end{equation}
Since $\underline I_d$ cannot exceed the exact sample mean $\bar I_d$, an approximation error not exceeding $\tau_{\mathrm a}+\tau_{\mathrm r}\bar I_d$, relative to exact DP on the same histories, is guaranteed upon acceptance. Otherwise, the threshold is evaluated by exact DP, in which the remaining incident coverage and the current shortage length are propagated hour by hour, so that any number of overlapping incidents is integrated exactly by Markov-chain imbedding~\cite{fu1994} (Appendix~\ref{app:dp}).

The procedure thus has two stages (Fig.~\ref{fig:overview}). In preparation, $c_d$, $b_d(s)$, and the distinct tuples $(d,z_t,m_t,o_t)$ with their multiplicities are stored for each history, which requires $O(T\sum_{d\in\mathcal D}d)$ operations for the thresholds $\mathcal D$. Each new scenario then requires $U(s)$ and $\Prb(D>s)$ for $s\le\max\mathcal D$, the weighted sum~\eqref{eq:firstorder}, and $O(J+J_{\mathrm h})$ operations for the local-bound terms, where $J$ is the number of distinct tuples and $J_{\mathrm h}$ their number summed over all histories, in addition to any exact fallback.

\subsection{Restoration Information Required at a Threshold}\label{sec:information}
The weighted sum~\eqref{eq:firstorder} depends on the restoration-time distribution only through $U(0),\ldots,U(d)$. Proposition~\ref{prop:information} shows that the exact expected exposure likewise depends only on the mean restoration time and the probabilities of restoration times below the threshold, none of which can be dropped.

\begin{proposition}[Restoration information]\label{prop:information}
For $\rho>0$ and $2\le d<T$, two bounded restoration-time distributions yield equal values of $\E_\theta[H_d\mid\omega]$ for every chronology of $A$, $B$, and $N$ hours if and only if they have equal means and equal probabilities $\Prb(D=j)$ for $1\le j<d$.
\end{proposition}

The proof is given in Appendix~\ref{app:information}. In practical terms, the probabilities of restoration times shorter than the threshold must be specified individually, whereas longer restoration times affect the expected exposure only through their total probability and their contribution to the mean.

\subsection{Estimation over Ordinary Histories}\label{sec:estimation}
Annual exposure is estimated from $n_0$ histories for the ordinary component and from a disjoint sample of $n_f$ histories for the fuel-induced increment. With exact increments, the estimator and its standard error are
\begin{equation}
\widehat\mu_d=\bar c_d+\bar I_d,\qquad
\widehat{\mathrm{SE}}=\left(\frac{\hat\sigma_c^2}{n_0}+\frac{\hat\sigma_I^2}{n_f}\right)^{1/2},
\label{eq:estimation}
\end{equation}
where $\hat\sigma_c^2$ and $\hat\sigma_I^2$ are the sample variances. The ordinary component is estimated once from a large sample and reused. Fuel-induced increments are evaluated on a smaller sample that is common to all scenarios, so that differences between scenarios are paired and their sampling variance is reduced. When the weighted sum is accepted, $\bar I_d$ is replaced by $\bar F_d$, and the approximation interval is reported with the standard error.

In retained-history MCS, incident paths are sampled on each history, so that the incident-sampling variance, divided by the number of paths per history, is added to the variance over histories, a term removed by conditional evaluation.

\begin{table*}[!t]
\caption{Test Cases}
\label{tab:design}
\centering\footnotesize
\setlength{\tabcolsep}{4pt}\renewcommand{\arraystretch}{1.0}
\begin{tabularx}{\textwidth}{@{}c l c r >{\raggedright\arraybackslash}X l l@{}}
\toprule
Case & Question & \shortstack{Demand\\multiplier} & Histories & Restoration distributions and incident rates & \shortstack[l]{Thresholds\\(h)} & Reference\\
\midrule
1 & Reliability, savings (\ref{sec:res-bound}, \ref{sec:res-cost}) & 1.20, 1.35 & 256 & 26 distributions, rates 0.025 to 1/year & 12, 24, 48, 96 & Reusable exact DP\\
2 & Statistical accuracy (\ref{sec:res-accuracy}) & 1.20 & 16 to 4096 & L at 0.05, 0.1, and 0.5/year & 24, 48 & Exact DP, MCS\\
3 & Planning insight (\ref{sec:res-restoration}) & 1.20 & 32\,768 & C, L, prevention, faster restoration, base rate 0.1/year & 24, 48 & Matched EUE, LOLH\\
\bottomrule
\end{tabularx}
\end{table*}

\section{Implementation Details}\label{sec:implementation}
\subsection{Test System and Data}
The Reliability Test System of the Grid Modernization Laboratory Consortium (RTS-GMLC)~\cite{barrows2020} was represented by its generator, transmission, and hourly profile data. The 2020 profiles were aligned to an 8760-hour year without February~29, regional demand was allocated to buses by the benchmark load shares, and hydro, wind, and photovoltaic profiles were imposed as hourly generation limits. Ordinary generator and branch availability was sampled from stationary two-state failure and repair models of the benchmark.

\subsection{Network Calculations and Evaluators}
For each ordinary history, hourly load shedding was minimized by a dc linear program solved with HiGHS~\cite{huangfu2018}, once with $\mathcal G$ at its sampled availability and once with $\mathcal G$ removed. Hours already determined by sufficient checks on island capacity deficits and feasible dispatch constructions were classified without optimization. Small violations of the ordering $\ell_t^0\le\ell_t^1$ caused by solver tolerances were corrected by $\ell_t^1\leftarrow\max(\ell_t^1,\ell_t^0)$, and shortages were identified with $\varepsilon=10^{-6}$~MW. In the exact DP, identical blocks were evaluated once, and the restoration tail beyond the threshold was replaced by two adjacent integer durations of equal total probability and mean, an exact replacement by Proposition~\ref{prop:information}.

\subsection{Numerical Verification and Computing Environment}
The evaluators were verified against an independent enumeration in 75-digit arithmetic over all three-class histories of up to five hours, five restoration-time distributions, and three incident rates. All 30\,060 expectations were enclosed by the bound within $5\times10^{-12}$~h, and the largest DP discrepancy in 18\,036 comparisons was below $9\times10^{-16}$~h. The sufficient network checks agreed with full optimization on all 1\,121\,280 endpoint-hours of 64 histories. The evaluators were implemented in Python with compiled numerical kernels and run serially on one core of an AMD Ryzen Threadripper 7980X workstation with 128~GiB of memory.

\section{Experimental Results}\label{sec:results}
The experimental evaluation is built around four questions, namely whether the local error bound is reliable (Section~\ref{sec:res-bound}), how much computation the method saves (Section~\ref{sec:res-cost}), whether statistical accuracy is preserved (Section~\ref{sec:res-accuracy}), and what the method reveals for planning (Section~\ref{sec:res-restoration}). The evidence comes from three cases on the RTS-GMLC benchmark, summarized in Table~\ref{tab:design} and described in Section~\ref{sec:res-setup}. In brief, no accepted result exceeded the requested tolerance, accepted scenarios were typically evaluated more than 200 times faster than by exact DP, the estimation error was nearly identical to that of exact DP, and alternatives with equal EUE and LOLH differed substantially in prolonged-shortage exposure.

\subsection{Experimental Design}\label{sec:res-setup}
The fuel-dependent group $\mathcal G$ comprised the ten 355-MW combined-cycle units and the twenty-seven 55-MW combustion-turbine units of RTS-GMLC, 37 units with a total capacity of 5035~MW. Hourly demand was scaled by factors of 1.20 and 1.35 to obtain two levels of dependence on this group. At these levels, ordinary LOLH of 12.2 and 192~h/year was obtained from 32\,768 and 256 histories, respectively, and supply was inadequate in 58\% and 73\% of hours after removal of the group. The median longest blocks of non-$N$ hours were 384 and 1148~h. These stressed levels produce long blocks, in which interactions between incidents are most frequent and error control is most demanding.

Two restoration-time distributions with a common mean of 39~h were used to isolate the effect of duration variability. Under the concentrated distribution C, restoration occurred after 36 or 42~h with equal probability. Under the extended-tail distribution L, durations of 12, 36, and 144~h occurred with probabilities $7/16$, $7/16$, and $1/8$. The tolerances in~\eqref{eq:acceptance} were set to $\tau_{\mathrm a}=10^{-4}$~h/year and $\tau_{\mathrm r}=1\%$.

Case 1 represents a planning study in which many interruption scenarios are reassessed on the same histories. Its 26 restoration distributions comprised C, L, and 24 lognormal and Weibull distributions with means of 24, 48, and 72~h and parent coefficients of variation of 0.25, 0.5, 1, and 1.5, which were discretized, bounded at 240~h, and adjusted to their means. Each distribution was evaluated at incident rates of 0.025, 0.05, 0.1, 0.2, 0.5, and 1/year, that is, from one shared fuel interruption every 40 years on average to one per year, giving 156 scenarios, and all distributions were evaluated at one rate before the next. The scenarios were assessed on 256 histories at thresholds of 24 and 48~h and, in an extended run, of 12, 24, 48, and 96~h. The proposed method and reusable exact DP were each run five times with fresh preparation and balanced order.

Case 2 tests whether statistical accuracy is preserved. The fuel-induced increment under distribution L at 0.05, 0.1, and 0.5/year was estimated twenty times from independent samples of 16 to 4096 histories by the proposed method, exact DP, and retained-history MCS with 64 or 256 paths per history. The estimates were compared by the relative root mean square error (RMSE) with an exact-DP reference from 32\,768 separate histories, whose standard errors at 0.1/year were below 0.03\% of the reference increments.

Case 3 compares mitigation alternatives with matched EUE and LOLH at demand 1.20 and a base rate of 0.1 incident/year, using exact increments from 32\,768 histories. In the first comparison, distributions C and L were compared, whose equal means give equal EUE and LOLH. In the second comparison, a frequency-reducing and a duration-reducing measure were represented in stylized form. A base distribution with durations of 12, 36, and 72~h, probabilities $1/4$, $1/2$, and $1/4$, and a rate of 0.1/year was modified by halving the rate (prevention) or every restoration time (faster restoration), which gives equal $\rho\E D$ and hence equal EUE and LOLH. Backup fuel that bridges the first $\delta$ hours of each interruption for the whole group is represented exactly by discarding incidents that end within $\delta$ hours and shortening the rest by $\delta$ hours, which replaces $\rho U(s)$ by $\rho U(s+\delta)$, so that no new network calculation is required.

\subsection{Reliability of the Error Bound}\label{sec:res-bound}
A reassessment can be trusted only if the bound never understates the actual error, and it saves time only if the bound is tight enough to accept most scenarios. Both properties were examined on all 1248 pairs of scenario and threshold in Case 1, its 156 scenarios at two demand levels and four thresholds, by comparing every first-order result with exact DP. Table~\ref{tab:acceptance} classifies the pairs at each threshold. A pair is accepted if the weighted sum passed the test~\eqref{eq:acceptance}. Otherwise, exact DP was applied, either because the bound was inconclusive although the actual error was within the tolerance $\tau_{\mathrm a}+\tau_{\mathrm r}\bar I_d$, or because the first-order value itself exceeded this tolerance. The last column measures the tightness of the bound by the median ratio of the bound to the actual error.

\begin{table}[!t]
\caption{Acceptance and Tightness of the Local Bound in Case 1}
\label{tab:acceptance}
\centering\footnotesize
\setlength{\tabcolsep}{4pt}
\renewcommand{\arraystretch}{1.0}
\begin{tabular}{@{}c r r r c@{}}
\toprule
& & \multicolumn{2}{c}{Exact DP applied} & \\
\cmidrule(lr){3-4}
\shortstack{Threshold\\(h)} & Accepted & \shortstack[r]{Bound\\inconclusive} & \shortstack[r]{First order\\inaccurate} & \shortstack{Median bound to\\actual error}\\
\midrule
12 & 312 & 0 & 0 & 1.02\\
24 & 306 & 6 & 0 & 1.09\\
48 & 207 & 81 & 24 & 17.3\\
96 & 63 & 210 & 39 & 52.2\\
\midrule
All & \textbf{888} & 297 & 63 & \\
\bottomrule
\end{tabular}
\par\smallskip
\begin{minipage}{\linewidth}\footnotesize
312 pairs per threshold. No accepted pair exceeded the tolerance.
\end{minipage}
\end{table}

The bound never fell below the actual error, and no accepted result exceeded the tolerance. At 12 and 24~h, the bound exceeded the actual error by a median of only 2\% and 9\%, and the weighted sum was accepted in 618 of 624 pairs. At longer thresholds, interactions between incidents become more important, the bound becomes more cautious, and exact evaluation was applied more often. Of the 360 fallbacks, only 63 were required because the first-order value itself was inaccurate, all at 48 or 96~h. At 48~h, these involved distributions with little probability of reaching the threshold, for which long runs arise mainly from overlapping incidents that the first-order sum omits.

The local bound was also compared with the block bound on the 24 pairs of distributions C and L at 0.1, 0.5, and 1/year and at 24 and 48~h. It was a median of 54 times narrower, with a range of 5.5 to 493, and accepted 14 of these pairs, whereas the block bound accepted none. The local bound thus permits acceptance where the block bound cannot, without ever accepting an inaccurate result.

\subsection{Computational Savings}\label{sec:res-cost}
Because exact fallback is the only expensive step after preparation, the savings follow directly from the acceptance rates. For each incident rate $\lambda$ of Case 1 and each demand multiplier, Table~\ref{tab:speedup} gives the share of scenarios accepted at both 24 and 48~h and the speed-up, the ratio of exact-DP time to proposed-method time. The last two rows give the speed-up of the complete workload of 156 scenarios at 24 and 48~h and at all four thresholds, including preparation and exact fallback.

\begin{table}[!t]
\caption{Speed-Up of the Proposed Method over Exact DP in Case 1}
\label{tab:speedup}
\centering\footnotesize
\setlength{\tabcolsep}{4pt}
\renewcommand{\arraystretch}{1.0}
\begin{tabular}{@{}l c c c c@{}}
\toprule
& \multicolumn{2}{c}{Scenarios accepted (\%)} & \multicolumn{2}{c}{Speed-up}\\
\cmidrule(lr){2-3}\cmidrule(lr){4-5}
Incident rate $\lambda$ (1/year) & 1.20 & 1.35 & 1.20 & 1.35\\
\midrule
0.025 & 96 & 88 & \textbf{23.8} & \textbf{13.5}\\
0.05 & 81 & 81 & 7.8 & 8.0\\
0.1 & 81 & 81 & 7.9 & 8.2\\
0.2 & 69 & 69 & 4.2 & 4.4\\
0.5 & 42 & 46 & 2.0 & 2.2\\
1 & 31 & 31 & 1.5 & 1.5\\
\midrule
All rates, 24 and 48~h & 67 & 66 & 3.27 & 3.56\\
All rates, 12 to 96~h & 21 & 20 & 1.20 & 1.32\\
\bottomrule
\end{tabular}
\par\smallskip
\begin{minipage}{\linewidth}\footnotesize
1.20 and 1.35 are demand multipliers. Speed-up is the median over five repetitions, using evaluation times in rate rows and total times in the last two rows.
\end{minipage}
\end{table}

Each accepted scenario was evaluated in a median of 1.1~ms at demand 1.20 and 2.8~ms at 1.35, against 266 and 1202~ms by exact DP, which is more than 200 times faster. The overall speed-up therefore follows the share of accepted scenarios. For rare incidents, at 0.025 to 0.1/year, 81\% to 96\% of the scenarios were accepted and, after preparation, the reassessment was 7.8 to 23.8 times faster. As the rate increases, overlapping incidents become more frequent, more scenarios need exact evaluation, and the speed-up falls to 1.5 at 1/year. The method is therefore most effective when incidents are rare.

Over the complete workload of 156 scenarios at 24 and 48~h, the total time, including preparation and exact fallback, was 3.27 and 3.56 times shorter, and the one-time preparation of 0.80 and 1.17~s was recovered by the fourth and the third scenario. With all four thresholds, only about a fifth of the scenarios were accepted at every threshold, and the speed-up decreased to 1.20 and 1.32. The proposed method thus remained faster in every tested workload, at the cost of retaining 12 to 46~MiB against 3 to 19~MiB for exact DP.

\subsection{Statistical Accuracy}\label{sec:res-accuracy}
\begin{figure}[!t]
\centering
\includegraphics[width=0.82\columnwidth]{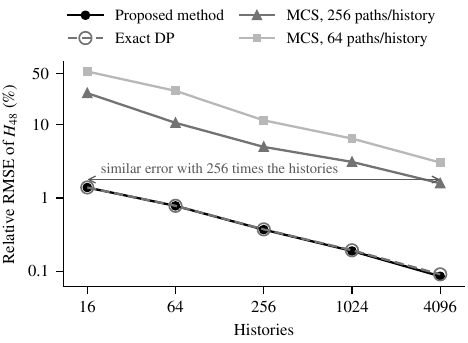}
\caption{Relative RMSE of the 48-h fuel-induced increment in Case 2 at 0.1 incident/year against the number of histories, over 20 assessments. The proposed method and exact DP coincide, as they also do at 24~h (Table~\ref{tab:accuracy}).}
\label{fig:convergence}
\end{figure}

An approximation that meets the tolerance on the retained histories could still increase the sampling error of the estimate. Fig.~\ref{fig:convergence} shows the relative RMSE of the 48-h fuel-induced increment in Case 2 at 0.1/year, and Table~\ref{tab:accuracy} gives the errors at both thresholds for selected sample sizes.

\begin{table}[!t]
\caption{Estimation Accuracy in Case 2}
\label{tab:accuracy}
\centering\footnotesize
\setlength{\tabcolsep}{5pt}
\renewcommand{\arraystretch}{1.0}
\begin{tabular}{@{}l r l c c@{}}
\toprule
& & & \multicolumn{2}{c}{Rel.\ RMSE (\%)}\\
\cmidrule(lr){4-5}
Method & Histories & Incidents & $H_{24}$ & $H_{48}$\\
\midrule
Exact DP & 16 & integrated & 1.34 & 1.39\\
\textbf{Proposed method} & 16 & integrated & \textbf{1.35} & \textbf{1.39}\\
MCS & 256 & 256 paths & 3.25 & 4.96\\
MCS & 4096 & 256 paths & 0.94 & 1.58\\
\bottomrule
\end{tabular}
\par\smallskip
\begin{minipage}{\linewidth}\footnotesize
At 0.1 incident/year under distribution L, over 20 independent assessments. MCS samples 256 incident paths per history.
\end{minipage}
\end{table}

The relative RMSE of the proposed method differed from that of exact DP by less than 0.05 percentage points at all rates and sample sizes, and the two curves coincide in Fig.~\ref{fig:convergence}. The accepted approximations therefore added only negligible error. Because both conditional methods integrate the incidents analytically, 16 histories sufficed for a relative RMSE of about 1.4\% at both thresholds. MCS samples the incidents instead, and with 256 paths per history, the smallest tested sample that stayed below 2\% at both thresholds comprised 4096 histories, each of which requires its own network calculations. For a single assessment, the evaluation by the proposed method, including preparation, took 54~ms against 46~ms for exact DP, and its savings arise in reassessment (Section~\ref{sec:res-cost}).

\subsection{Planning Insight}\label{sec:res-restoration}
\begin{figure}[!t]
\centering
\includegraphics[width=0.82\columnwidth]{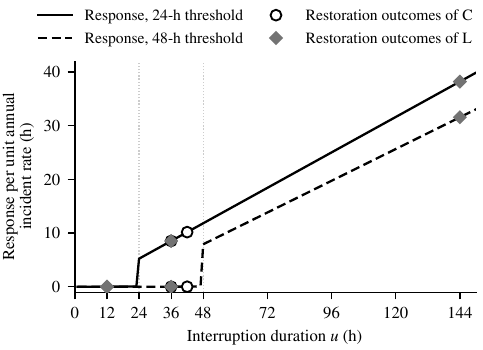}
\caption{Restoration response at demand 1.20, $\Psi_d(u)/8760$, which is the first-order increase of the expected exposure per unit annual incident rate when every incident lasts $u$ hours, averaged over 32\,768 histories. Markers place the outcomes of C and L on each response, and the 36-h outcome is shared.}
\label{fig:response}
\end{figure}

The last question is what the method reveals that EUE and LOLH do not. Table~\ref{tab:application} gives the exposures of the two comparisons of Case 3, and Fig.~\ref{fig:response} explains them through the restoration response.

\begin{table}[!t]
\caption{Prolonged-Shortage Exposure in Case 3}
\label{tab:application}
\centering\footnotesize
\setlength{\tabcolsep}{6pt}
\renewcommand{\arraystretch}{1.0}
\begin{tabular}{@{}l l >{\centering\arraybackslash}p{1.25cm} >{\centering\arraybackslash}p{1.25cm}@{}}
\toprule
& & \multicolumn{2}{c}{Exposure ($10^{-2}$~h/year)}\\
\cmidrule(lr){3-4}
Comparison & Alternative & $H_{24}$ & $H_{48}$\\
\midrule
Same mean duration & Distribution C & 99.3 & \textbf{0.4}\\
& Distribution L & \textbf{90.9} & 39.8\\
\addlinespace
Same unavailability & Prevention & 50.2 & 17.6\\
& Faster restoration & \textbf{27.2} & \textbf{0.3}\\
\bottomrule
\end{tabular}
\par\smallskip
\begin{minipage}{\linewidth}\footnotesize
Within each comparison, both alternatives give equal EUE and LOLH, and the lower exposure is in bold. Exact increments from 32\,768 histories plus an ordinary baseline from 245\,760 disjoint histories. Standard errors are 0.28 at 24~h and 0.08 at 48~h, dominated by the baseline, which cancels from paired differences.
\end{minipage}
\end{table}

In the first comparison, both distributions gave an LOLH of 14.5~h/year, yet replacing C by L reduced the 24-h exposure but increased the 48-h exposure, with paired changes of $(-8.431\pm0.003)\times10^{-2}$ and $(39.357\pm0.010)\times10^{-2}$~h/year ($\pm$ one standard error). Fig.~\ref{fig:response} shows why. The response is nearly zero below the threshold and jumps when the duration reaches it, because a shorter incident can complete a prolonged shortage only together with ordinary shortages. Both outcomes of C, 36 and 42~h, lie below 48~h and contribute almost nothing to the 48-h exposure, whereas the 144-h outcome of L contributes almost all of it. At 24~h, the 12-h outcome of L contributes almost nothing, which lowers its exposure below that of C. Both changes were established without exact integration, by response intervals of $[-0.0846,-0.0839]$ and $[0.3919,0.3954]$~h/year.

In the second comparison, faster restoration reduced the 24-h exposure to 0.272~h/year, against 0.502~h/year under prevention, a reduction of approximately 46\%, because halving every duration moves most outcomes below the threshold, where the response is nearly zero. This ranking is not universal. Faster restoration spreads the same fuel unavailability over twice as many shorter incidents, which rarely cover two distant hours together. Where the coverage of either of two such hours already causes a prolonged shortage, this spreading creates more shortages, and prevention becomes preferable, as the chronology of Appendix~\ref{app:reversal} shows.

Two implications for planning follow. Prolonged-shortage exposure should be evaluated alongside EUE and LOLH, because equal values of the latter can conceal large differences and even opposite rankings, in line with earlier findings that expectation-based metrics obscure important features of shortfall events~\cite{stenclik2023,leibowicz2024}. The ranking of measures depends on the system and must be evaluated for it, which the proposed method makes inexpensive.

\section{Conclusion}\label{sec:conclusion}
A reassessment method for prolonged-shortage exposure under shared fuel interruptions has been presented, in which network calculations are performed once and condensed into a restoration response, and exact DP is applied only when a local error bound cannot accept the resulting weighted sum.

The experiments on the RTS-GMLC benchmark support four conclusions. First, the bound is reliable, because no accepted result exceeded the tolerance in 1248 tested pairs of scenario and threshold, and its median overestimate of the actual error was at most 9\% at thresholds up to 24~h. Second, the method saves most computation when incidents are rare, because accepted scenarios were typically evaluated more than 200 times faster than by exact DP and, after preparation, rare incidents were reassessed up to 24 times faster. Third, accuracy is preserved, because the relative RMSE differed from that of exact DP by less than 0.05 percentage points, and 16 histories gave an error similar to that of MCS with 4096 histories and 256 paths per history. Fourth, EUE and LOLH are insufficient for comparing measures against fuel interruptions, because alternatives with equal values of both differed substantially in prolonged-shortage exposure. In a stylized comparison, faster restoration gave 46\% less 24-h exposure than prevention, and the ranking of alternatives could reverse between thresholds.

Measures against shared fuel interruptions can therefore be screened quickly by their effect on prolonged shortages, with an analytical error bound keeping every result within a requested tolerance of exact evaluation on the same histories. Since the exact exposure depends on restoration times only through their mean and their probabilities below the threshold (Proposition~\ref{prop:information}), restoration data need to be characterized in detail only below the threshold of interest.

The method is formulated for incidents that occur independently of system conditions at a stationary rate, a fixed generator group that is removed as a whole, and hourly dispatch without time coupling. Extensions to interruptions that coincide with high demand, as in cold-weather events, and to seasonal incident rates, partial fuel curtailment, and time-coupled operation such as energy storage and unit commitment are left for future work. The cost of fallback could also be reduced by applying exact evaluation only to the hours that make the bound inconclusive.

\appendices
\section{Proof of Proposition 1}\label{app:response}
If the first and last required hours of an eligible window are $f$ and $l$, all of them are covered by one incident of duration $u$ if and only if the incident is started between $l-u+1$ and $f$, that is, at one of $(u-s)_+$ integer hours, where $s=l-f$. Summation of~\eqref{eq:window} over the starts yields~\eqref{eq:response}. Windows without required hours are unaffected by the incident and cancel in $c_d$.

\section{Block Bound}\label{app:block}
Over blocks $\beta$ of consecutive non-$N$ hours, the remainder satisfies
\begin{equation}
|R_d|\le\kappa_d^{\mathrm{blk}}=\sum_\beta\Gamma_{\beta,d}\,\varphi_1\big(\rho(|\beta|-1+\E D)\big),
\label{eq:blockbound}
\end{equation}
where $\Gamma_{\beta,d}=|\beta|\ind\{|\beta|\ge d\}-H_d(\ind_{A\cap\beta})$ is the largest additional reward in $\beta$ and $\varphi_1(x)=x-1+e^{-x}$ is the expected Poisson count beyond the first event for mean $x$. Within $\beta$, the additional reward is nonnegative, nondecreasing in coverage, and bounded by $\Gamma_{\beta,d}$, so that the combined increment of $q$ incidents differs from the sum of their individual increments by at most $(q-1)_+\Gamma_{\beta,d}$. The number of incidents intersecting $\beta$ is Poisson-distributed with mean $\rho(|\beta|-1+\E D)$, and the expected count beyond the first equals $\varphi_1$ of this mean, which yields~\eqref{eq:blockbound}. Since $\varphi_1(x)\le x^2/2$ and $\Gamma_{\beta,d}\le|\beta|$, a block contributes at most $\rho^2|\beta|(|\beta|-1+\E D)^2/2$, which is cubic in the block length.

\section{Local Error Bound}\label{app:local}
Membership of $t\in\Qd$ in a qualifying run is indicated by $g_t$, so that
\begin{equation}
H_d-c_d=\sum_{t\in\Qd}g_t.
\label{eq:membership}
\end{equation}
For $t\in\Qd$, every eligible $d$-hour window containing $t$ contains a required hour, and these windows have consecutive starts because they are bounded by class-$N$ hours. All required hours of window $w$ are covered by one incident of duration $u$ for starts in $[l_w-u+1,f_w]$. The number of starts for which $t$ is made to qualify by one incident of duration $u$ is therefore
\begin{equation}
\xi_t(u)=\Big|\bigcup\nolimits_w\big([l_w-u+1,f_w]\cap\mathbb Z\big)\Big|,
\label{eq:kt}
\end{equation}
and the intensity of incidents by which $t$ is individually made to qualify is $\alpha_t=\rho\E\xi_t(D)$, with $\sum_{t\in\Qd}\alpha_t=F_d$. The largest of these intervals contains $(u-m_t)_+$ integers, and all of them are contained in the integer envelope $[l_{\mathrm{first}}-u+1,f_{\mathrm{last}}]$, which contains $(u-o_t)_+$ integers. Since $l_w\ge l_{\mathrm{first}}$ and $f_w\le f_{\mathrm{last}}$ for every window, $m_t\ge o_t$. For $u\ge d$, consecutive intervals overlap or are adjacent on the integer lattice, because $l_{w+1}-f_w\le d\le u$, so that their union equals the envelope. Hence
\begin{equation}
\begin{aligned}
\xi_t(u)&\ge(u-m_t)_+ +(m_t-o_t)\ind\{u\ge d\},\\
\xi_t(u)&\le(u-m_t)_+ +(m_t-o_t)\ind\{u>m_t\}.
\end{aligned}
\label{eq:ktbounds}
\end{equation}
Taking expectations over $D$ gives $\alpha_t^-\le\alpha_t\le\alpha_t^+$, with $\alpha_t^\pm$ as in~\eqref{eq:alpha}. An incident of duration $u$ intersects the hours from $f_{\mathrm{first}}$ to $l_{\mathrm{last}}$ for $u+z_t$ start times, so that the number of incidents that can affect $g_t$ is Poisson-distributed with mean $\nu_t$.
\begin{IEEEproof}[Proof of Proposition~\ref{prop:local}]
By Poisson thinning, the number of individually sufficient incidents is Poisson-distributed with mean $\alpha_t$, and qualification of $t$ is ensured by at least one of them, because additional coverage cannot shorten a run. Conversely, qualification is possible only if at least two relevant incidents occur, or if the only relevant incident is individually sufficient. Hence
\begin{equation}
1-e^{-\alpha_t}\le\E g_t\le\varphi_2(\nu_t)+\alpha_te^{-\nu_t}.
\label{eq:membershipbound}
\end{equation}
After subtraction of $\alpha_t$, both sides are nonincreasing in $\alpha_t$. Substitution of~\eqref{eq:alpha} and summation over $t$ by~\eqref{eq:membership}, with $\sum_t\alpha_t=F_d$, yields~\eqref{eq:localbound}. Since $\alpha_t^+\le\nu_t$, $\varphi_2\le\varphi_1$, $\varphi_1(x)\le x^2/2$, and $z_t\le2d-2$, both bounds are at most $\frac{\rho^2}{2}\sum_{t\in\Qd}(\E D+z_t)^2$ in magnitude, which gives~\eqref{eq:scaling}.
\end{IEEEproof}

\section{Exact Dynamic Program}\label{app:dp}
In the DP, the remaining incident coverage and the length of the current shortage run, capped at $d$, are propagated hour by hour. With $M_t$ denoting the longest duration among incidents started at hour $t$ (zero if none) and $X_t$ the largest remaining coverage including hour $t$,
\begin{equation}
\begin{aligned}
\Prb(M_t\le x)&=e^{-\rho\Prb(D>x)},\\
X_t&=\max\{(X_{t-1}-1)_+,M_t\},
\end{aligned}
\label{eq:dp}
\end{equation}
with stationary initialization $\Prb(X_{-1}\le x)=e^{-\rho U(x)}$. Hour $t$ is short if $t\in A$, or if $t\in B$ and $X_t>0$. When a run first reaches $d$ hours, $d$ hours are credited, and one further hour is credited for each continuation. The run length is initialized at zero to implement the clipped reward. Let $\gamma_{a,r}$ denote the probability of remaining coverage $a$ and run length $r$ after the coverage has been decremented, with coverage zero and one combined at $a=0$. With $\Phi(j)=e^{-\rho\Prb(D>j)}$, $\Phi(-1)=0$, and $p_j=\Phi(j)-\Phi(j-1)$, the coverage is updated as
\begin{equation}
\widetilde\gamma_{j,r}=p_j\sum_{a=0}^{j}\gamma_{a,r}+\Phi(j-1)\gamma_{j,r}.
\label{eq:dptransition}
\end{equation}
The first term represents coverage set by a new incident, and the second longer coverage remaining from an earlier one. The run length is then increased to $\min(r+1,d)$ on a shortage hour and reset to zero otherwise. Blocks separated by class-$N$ hours are evaluated separately from stationary coverage marginals.

\section{Proof of Proposition 3}\label{app:information}
The probability that hours $t_1<\cdots<t_K$ are all uncovered is
\begin{equation}
\Prb(\chi_{t_1}=\cdots=\chi_{t_K}=0)
=e^{-\rho[KU(0)-\sum_{i=2}^{K}U(t_i-t_{i-1})]}.
\label{eq:void}
\end{equation}
Since every required set in~\eqref{eq:window} has a span of at most $d$, its coverage probability is obtained from $U(0),\ldots,U(d)$ by inclusion and exclusion, which establishes sufficiency. For necessity, blocks with one or two required hours, bounded by class-$N$ hours and otherwise in class $A$, are used. With $\pi=\Prb(\chi_0=1)$ and $\pi_s=\Prb(\chi_0=\chi_s=1)$, a reward of $d\pi$ is obtained for one required hour in a length-$d$ block. In a length-$(d+1)$ block, a reward of $d\pi+\pi_s$ is obtained for required positions $0$ and $s$ with $1\le s<d$, and $2d\pi-(d-1)\pi_d$ for required endpoints. Equality for every chronology therefore identifies $\pi$ and $\pi_1,\ldots,\pi_d$, and, through
\begin{equation}
\pi=1-e^{-\rho U(0)},\quad
\pi_s=1-2e^{-\rho U(0)}+e^{-\rho[2U(0)-U(s)]},
\label{eq:pairs}
\end{equation}
the values $U(0),\ldots,U(d)$. Necessity follows from $U(0)=\E D$ and $\Prb(D=j)=U(j-1)-2U(j)+U(j+1)$. Conversely, $U(0),\ldots,U(d)$ are determined by the mean and probabilities below $d$.

\section{Example of a Reversed Ranking}\label{app:reversal}
This appendix constructs a chronology in which faster restoration, unlike in Section~\ref{sec:res-restoration}, yields a larger 24-h exposure than prevention at equal EUE and LOLH.

Consider a block of 25 hours whose endpoints, hours 0 and 24, are in class $B$ and whose 23 interior hours are in class $A$, bounded by class-$N$ hours. Coverage of at least one endpoint produces a run of at least 24~h, so that $\E H_{24}=48\pi-23\pi_{24}$, where $\pi$ is the probability that an hour is covered and $\pi_{24}$ that of both endpoints being covered. An endpoint is uncovered with probability $e^{-\rho\E D}$. Since $\rho U(24)$ is the expected number of incidents covering both endpoints, the number of incidents covering at least one of them is Poisson-distributed with mean $\rho(2\E D-U(24))$. By inclusion and exclusion,
\begin{equation}
\pi_{24}=1-2e^{-\rho\E D}+e^{-\rho(2\E D-U(24))}.
\label{eq:pairpi}
\end{equation}
The base distribution of Section~\ref{sec:res-setup} has $\E D=39$~h and $U(24)=18$~h at the hourly rate $\rho$. Prevention halves the rate, which gives the exponents $19.5\rho$ and $30\rho$ in~\eqref{eq:pairpi}. Faster restoration halves every duration, so that $\E D=19.5$~h and $U(24)=3$~h at the full rate, which gives $19.5\rho$ and $36\rho$. Both measures therefore give the same $\pi$, and faster restoration gives the larger exposure, by $23(e^{-30\rho}-e^{-36\rho})$.

\section*{Acknowledgment}
The author used Claude (Anthropic) to edit the manuscript for language, clarity, and readability. The author reviewed and edited the output and takes full responsibility for the content of this publication.

The author declares no known competing financial interests or personal relationships that could have appeared to influence the work reported in this paper.

\end{document}